\documentclass[sigconf, anonymous=false]{acmart}
\usepackage{booktabs} 

\usepackage{graphicx}
\usepackage{subcaption}
\usepackage{amssymb}
\usepackage{amsmath}
\usepackage{amsfonts}
\newtheorem{defn}{Definition}

\acmConference[ ]{ }{}{} 
\acmYear{2018}
\copyrightyear{2018}

\begin{document}

\title{A Social Network Analysis Framework for\\Modeling Health Insurance Claims Data}

\author{Ana Paula Appel} 
\affiliation{
   \institution{IBM Research\\}}
\email{apappel@br.ibm.com}

\author{Vagner F. de Santana} 
\affiliation{
   \institution{IBM Research\\}}
\email{vagsant@br.ibm.com}

\author{Luis G. Moyano}
\affiliation{ 
   \institution{FCEN/CONICET, Mendoza, Argentina}}
\email{lgmoyano@mendoza-conicet.gob.ar}

\author{Marcia Ito} 
\affiliation{
   \institution{IBM Research\\}}
\email{ marciaito@br.ibm.com}

\author{Claudio Santos Pinhanez}
\affiliation{
   \institution{IBM Research\\}}
    \email{csantosp@br.ibm.com}

\begin{abstract}
  Health insurance companies in Brazil have their data about claims organized having the view only for providers. In this way, they loose the physician view and how they share patients. Partnership between physicians can view as a fruitful work in most of the cases but sometimes this could be a problem for health insurance companies and patients, for example a recommendation to visit another physician only because they work in same clinic. 
  The focus of the work is 
  to better understand physicians activities and how these
  activities are represented in the data.
  Our approach considers three
  aspects: the relationships among physicians, the relationships between physicians and patients, and
  the relationships between physicians and health providers.  
  We present the results of an analysis of a claims database (detailing 18
  months of activity) from a large health insurance company in Brazil.  
  The main contribution presented in
  this paper is a set of models to represent: \emph{mutual referral} between physicians, 
  \emph{patient retention}, and \emph{physician centrality} in the health insurance network. Our results show the proposed models 
  based on social network frameworks, extracted surprising insights about
  physicians from real health insurance claims data.
\end{abstract}

\sloppy




\maketitle

\section{Introduction}

Health insurance companies have an important piece of transactional data in their ecosystem that are the claims. A claim represents a
report from a physician or a healthcare service provider to a health insurance company, requesting some form of fee related to a patient's consultation with a physician, a clinical exam, or a medical procedure.  Even though claims data may vary, it
generally contains at least the ID of the healthcare professional involved in the procedure (it may also be a group of professionals), the ID of the patient, the type of procedure, and time information related to the event. It may include other types of information such as location of the service, CID, cost, cost payed, pre-authorization codes, etc.
%

Traditionally the analysis of claims data is based on applying statistics to the individual elements of the system (physicians, service providers, patients) or to the set of claims in order to produce reports of cost or quantity of procedures. With the system organized only to pay providers that could be physicians, hospitals or clinics the health insurance company looses almost for complete the view of physician in this system. However, for a health insurance company is very important to know how physicians relationship among themselves, how patiences flow from one physician to another and how the patients are coming back to the physician and if all this are compliance with health insurance company. 



In practice, get the relation among physicians using claims is difficult because claims are paid to a wide variety of providers, hospitals, clinics, or even physicians registered as small companies. A single physician may contact a patient through all those channels. 
Also, there is a large variety of services that a health insurance company offer combined with the problem that in Brazil a patient can go direct to an specialist. She/He does not need to go in a general physician before reach a cardiologist or a immunologist.  
In spite of all those difficulties, we show in this paper that meaningful and reliable models, based on social network frameworks, about the relationships about physicians and patients can be computed from claim data.





This project was a short term project develop with a large health insurance company in Brazil. The main contribution of this work is a set of models to identify relationship-based patterns related to physicians excellence or, on the other hand, possible abusive practices. These models can therefore be used by health insurance companies to better manage the physicians they have businesses with. It can also be used to support patients to receive more integrated care from a group of physicians and service providers. The proposed framework is composed of the following models: 

\begin{itemize}
\item \textbf{Mutual Referral}: a proxy for physicians that refer patients to each other.
\item \textbf{Retention}: summarizes how physicians retain patients over time, including returning behavior in patient-physician pairs.
\item \textbf{Physician Centrality}: summarizes that relative importance of physicians in the physicians-physicians network.
\end{itemize}

The context of this paper is the analysis of a large claims database from a major Brazilian
health insurance company. 
We use this database to identify patterns in physicians behaviors and then model their relationships by means of social network frameworks. The work on this database both inspired the development of the proposed models and also provide empirical validation for them.

This paper is organized as follows: section \ref{sec:background} describes the related work, section \ref{sec:data} details the database analyzed, section \ref{sec:metrics} presents the proposed models, 
and section \ref{sec:conc} concludes.

\section{Related Work} \label{sec:background}

Healthcare data is heralded as the key element in the quest to improve efficiency and
reduce costs in healthcare systems\cite{srinivasan2013leveraging}. This trend is becoming more pronounced as multiscale data generated from individuals is continuously increasing, particularly due to new high-throughput sequencing platforms, real-time imaging, and point-of-care devices, as well as wearable computing and mobile health technologies
\cite{andreu2015big}.

In healthcare, data heterogeneity and variety arise as a result of linking the diverse range of
biomedical data sources available. Sources of quantitative data (e.g., sensor data, images, gene
arrays, laboratory tests) and qualitative data (e.g., diagnostics, free text, demographics) usually
include both structured and unstructured data, normally under the name of Electronic Health Records.
Additionally, the possibility to process large volumes of both structured and unstructured medical
data allows for large-scale longitudinal studies, useful to capture trends and to propose
predictive models\cite{jensen2012mining}.

One of the most useful and commonly used datasets are claims databases. 
Claims data is often rich in details, as it describes important elements of the events taking place around
the healthcare professional and the patient, e.g., timestamps, geographical location, diagnosis
codes, associated expenses, among others.
The use of claims data in healthcare studies has been scrutinized in \cite{burton2011healthcare} and \cite{Chandola2013},
providing a set of good practices and outlining the shortcomings of claims-based research.

Social network framework has proven to be a useful analysis tool in this context, allowing for insights
difficult to reach by traditional descriptive statistics as presented in \cite{Chandola2013}.
%
%
For instance, social network analyses has been used to study comorbidity, the simultaneous presence of two chronic diseases or conditions in a patient.
%
%
By structuring diseases as a network, it is possible to quantify some of the aspects of the complex interactions between conditions in the different patient populations. 
A number of studies have focused on extensive claims datasets  to examine and understand comorbidity networks.
In \cite{chmiel2014spreading}, the authors study a diffusion process on a comorbidity network to
model the progressive spreading of diseases on a population depending on demographic data.
In \cite{klimek2015quantification}, the authors study how a given chronic disease (diabetes)
correlates with age and gender, spanning almost 2 million patients from an entire European country.
Such comorbidity networks have also been proposed as models to understand the connection between
genetic and environmental risk factors for individual diseases \cite{klimek2016disentangling}.

Beyond clinical purposes, claims data have also been studied to understand the complex interactions of
different organizational structures and management relationships involved in patient care processes.
For instance, in \cite{lee2011mining}, temporal patterns in Electronic Health Records were modeled
in order to present useful information for decision-making. The authors developed a data
representation for knowledge discovery so as to extract useful insights on latent factors
of the different processes involving a patient with the aim to improve workflows.
%

Another important trend is the capturing the relationship
between healthcare professionals, in particular the physicians.
In \cite{landon2013using}, the authors apply social network analyses to mine  
networks of physicians which might be used to improve the designation of middle-sized administrative
units (accountable care organizations).
Sauter et al. \cite{sauter2014analyzing} use social network to understand networks of healthcare providers
which share patients, giving insight in the interplay between general practitioners, internal
specialists, and pediatricians.
Also, the network structure of the different healthcare providers who care for a given individual can
show important variability of the healthcare system \cite{mandl2014provider}.

Social networks have also been used to understand the state of coordination of healthcare actors.  In
\cite{wang2014application} the authors  describe a complex network approach applied to health insurance claims to understand the nature of collaboration among physicians treating hospital in-patients and to explore the impact of collaboration on cost and quality of care. Also, in \cite{uddin2014social}, the authors study the social network structure in hospitals among healthcare professionals to understand which
variables affect patient care efficiency measures.
The idea is further developed in \cite{uddin2016exploring} from a statistical point of view in a
medium-sized number of hospitals, through the analysis of  temporal patterns and costs.

The medical referral system in the Canadian healthcare system is studied  in \cite{almansoori2011applications}, where the authors map and analyze the network between general practitioners and specialists . In \cite{guo2015find} the authors describe the condition of the basic medical insurance for urban
and rural residents in China, then they demonstrate that social network analysis can be used in the health insurance claims data to support better understanding of patients transfers among hospitals.

Finally, social network visualization methods can also be powerful to explore and analyze healthcare
information, in particular to depict the relationship among healthcare professionals \cite{moyano2016graphys}.

\section{The Insurance Claims Database} \label{sec:data}


The data used in this study was provided by a large Brazilian health insurance company. The database contains information about services and materials of 108,982,593 instances of claims paid by the health insurance company to service providers covering 18 months of activity, about 200,000 claims per day. 
The databased names 279,085 physicians, of which 81\% are considered valid, 
that is,  the physician register ID is well formed.
Moreover, we have information about 2,243,198 patients and 26,033 providers.  The claims data contain only claims related to medical consultations performed by physicians, and did not include claims related to clinical analysis, image-based exams, or hospitalizations.

Two important aspects related to the quality of the data regarding state information and physician specialties need to be mentioned. First, approximately 25\% of the data do not contain information about state in which the service was performed. This proportion increases when the data is modeled as a graph, since that missing state value in one or both nodes (representing physicians) may invalidate an important piece of the analysis. 
Another important aspect of the data is related to the distribution of physicians' specialties. This attribute is important to correlate with physicians' relationships and crucial to the proper understanding of the results. However, because of the large amount of missing values, the use of this information in the data analysis had limited scope.



\subsection{Mapping Claims Data as a Graph} \label{sec:graph}

As mentioned before, the dataset used contains only claims related to procedures performed by physicians. 
%
%
%



In this work we focus on the relationships between physicians within the health insurance company network only for consultations. Thus, two physicians are considered related if they have a common patient, that is, a patient that had a consultation with both physicians. This does not indicate a direct relationship, but, for large number of common patients (represented as an outlier), there is a high likelihood that these physicians have some kind of professional relationship, be it a similar profile, same provider, similar location, similar education background, etc. This signals the possibility that they know each other and have referred their  patients to one another.

In order to model a physician-physician network, we build a network model as a graph $G = (V, E)$, where the $\mathcal{\left|V\right|} = N$ denotes the set of nodes that represent physicians. $\mathcal{\left|E\right|}=M$ denotes the edges that connect physicians that have in common consultations claim of a certain patient, $e_k \in \mathcal{E}$ and $e_{ij} = \left\{(v_i,v_j)| v_i,v_j \in \mathcal{V} \right\}$. 

The advantage in using a social network to model the relationships among physicians involves the several ways to quantify the relative importance of the physicians.
For example, social network analysis allows for individualization of physicians that are in a prominent position in the network be it due to the relationship with her peers, due to her connections with influential physicians, or because without the physician the topology of the network would change substantially (e.g., by increasing the number of connected components). 



\begin{figure}[htb]
\centering
\includegraphics[width=0.44\textwidth]{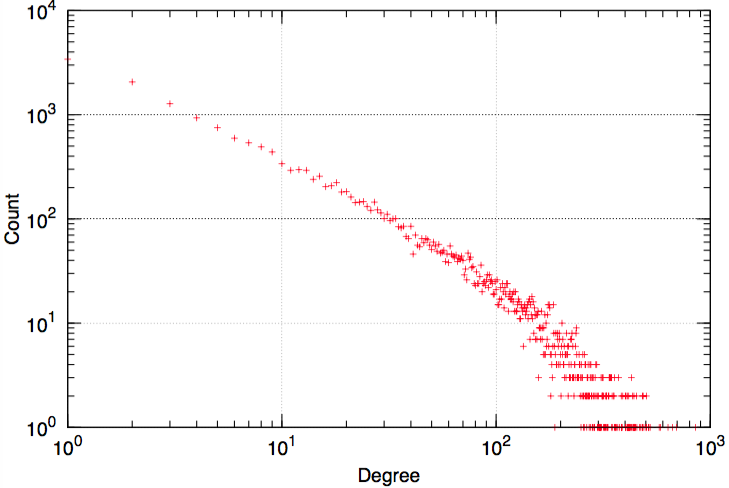}
\caption{Physician-physician network degree distribution.}
\label{fig:degree}
\end{figure}

Figure \ref{fig:degree} shows the degree distribution of the physician-physician network. As we can
see, the tail of the distribution follows approximately a power-law, pointing to the fact that the
relationship between physicians has a  structure also commonly found in other real networks, such as
social networks \cite{Newman2010}. Here, most physicians are connected with only a few other
physicians while a small number of physicians are very well connected in the network.

In the next sections we describe in detail the models proposed to better understand the physicians' behavior in this type of network.

\section{Modeling Claims Data} \label{sec:metrics}

This section details the proposed framework for modeling health insurance claims data. It considers healthcare service relationships from three different perspectives: (1) how physicians refer each other, (2) how physicians retain patients over time, and (3) how important physicians are in the physician-physician network.

\subsection{Mutual Referral}

Identifying physicians that work together, especially for consultations, is of great business value for health insurance companies. Physicians working together may be due to several reasons. On the one hand, it could be positive for medical care professionals to treat patients together and be able to work as a team, building trust. On the other hand, this behavior could also point to a misuse of health insurance resources. For instance, it could be the case that every time a patient visits a given clinic or physician, she is redirected to another physician, even if there is no need to do so. Other situations could involve physicians benefiting certain colleagues by issuing unnecessary referrals, possibly under terms of reciprocity.  


Accordingly, we set to define a model with the goal of identifying physicians that refer each other. 
The physician-physician network used to compute this metric is a slightly modified version of the network proposed in Section \ref{sec:graph}. Here we consider the consultations timestamps, i.e., two physicians $P_1$ and $P_2$ are connected if the same patient has a claim about consultation both with $P_1$ and with $P_2$ but edges retain their chronological order so as to indicate the direction of the link. As a result, we obtain a directed network, i.e., a patient consulting first $P_1$ and then $P_2$ generates and edge $e_{12}$ and a patient consulting $P_2$ in the first place and only afterwards $P_2$ generates an edge $e_{21}$.

Thus, we consider the physician-physician network $G(V,E,w)$ with node $v_i \in V$ and edges $e_{ij} \neq e_{ji} \in E$. Each edge $E$ has an associated weight $w: E \rightarrow \mathbb{R}^+$ equal to the number of patients that consulted with physician $P_1$ and subsequently with $P_2$, denoted as $w_{ij}$. 






\begin{defn}[Mutual Referral]
  The \textit{Mutual Referral} metric focuses on identifying pairs of nodes $v_i$ and $v_j$ (where $i, j \in [1, N]$ being $N$ the total number of nodes in the network $G$) connected by edges $e_{ij}$ and $e_{ji}$, where the weights $w_{ij}$ and $w_{ji}$ are high (the metric increases proportionally to the weights) and, at the same time, as similar as possible to each other (the metric decreases as weights differ):
\begin{equation*}
mr(v_i,v_j) = w_{ij} + w_{ji} - |w_{ij} - w_{ji}|,
\end{equation*}

\end{defn}

where $mr(v_i,v_j) = mr(v_j,v_i)$. Our metric is symmetric to $i, j$, as expected for a variable describing a property of a given pair of nodes. Note that the second term in the metric $mr$ represents a penalty for those pairs of edges not similar to each other, thus the metric scores higher in case of symmetrical relationships.




To allow for the possibility of a global comparison between pairs of nodes, we defined a mutual referral score ($mrs$), by normalizing by the maximum value of $mr$ across the complete graph $G$. 

\begin{defn}[Mutual Referral Score]
The \textit{Mutual Referral Score} measures, in the unit range $[0,1]$, the relationship between each pair of nodes $v_i$ and $v_j$, relative to the maximum mutual referral identified in the directed graph $G$, represented by $max(mr(v_s,v_t))$, $\forall s,t \in [1, N]$. 
Thus, for any pair of nodes $v_i$ and $v_j$ in $G$, the mutual referral score is defined as: 
\begin{equation*}
mrs(v_i,v_j)= \frac{mr(v_i,v_j)}{max(mr(v_s,v_t))},\  \forall s,t \in [1,N],
\end{equation*}

\end{defn}


where again $mrs(v_i,v_j)=mrs(v_i,v_j)$. 

Aiming at identifying connections among physicians when the studied settings change (e.g., population, availability of health services, socioeconomic levels), analyses were performed considering the top 20 and top 50 physician IDs with strongest mutual referral scores from five Brazilian states containing the most claims. 

\begin{figure*}[htbp]
\centering
\includegraphics[width=0.9\textwidth]{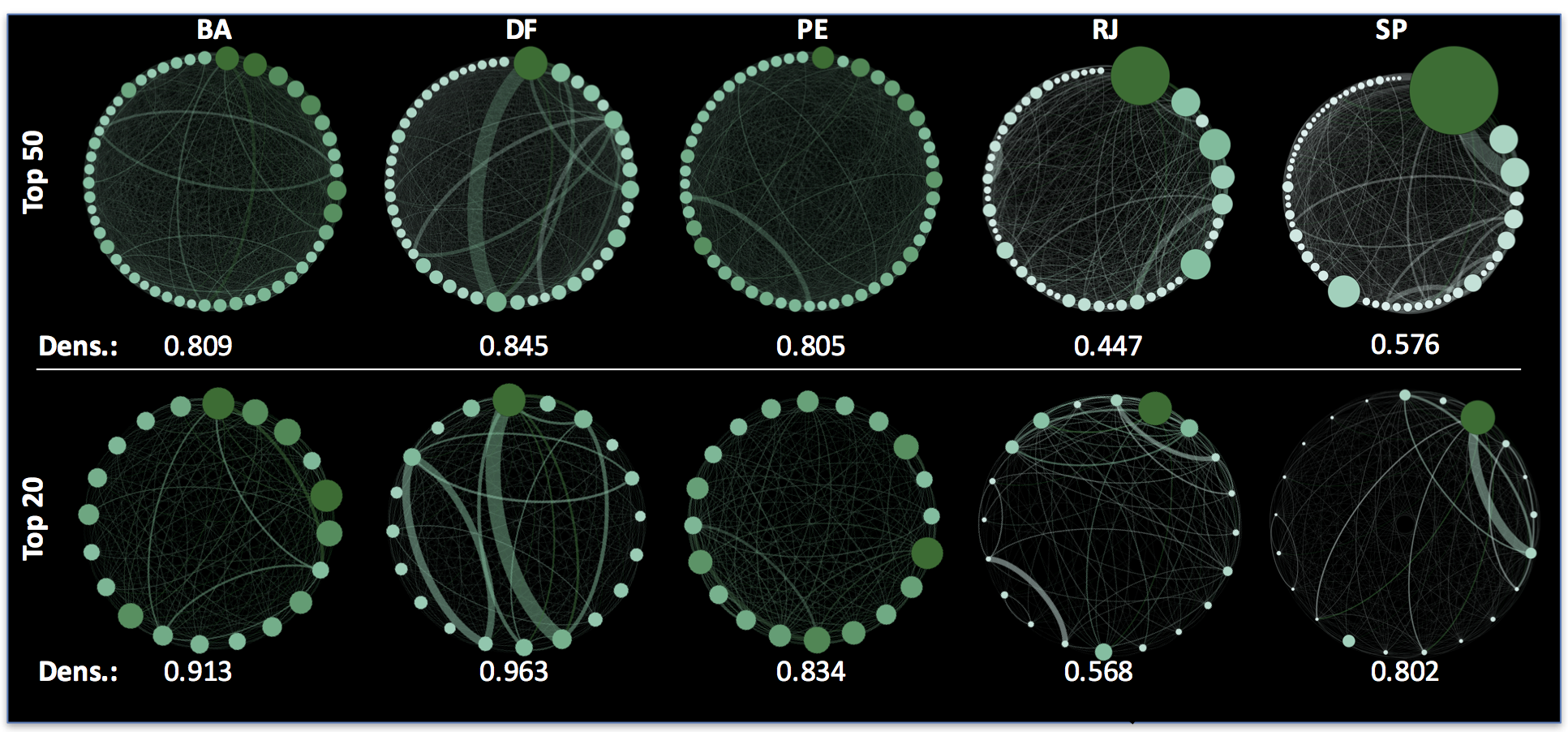}
\caption{Top 50 and 20 physicians with highest mutual referral score from the following Brazilian states: Bahia (BA), Pernambuco (PE), Distrito Federal (DF), S\~ao Paulo (SP) and Rio de Janeiro (RJ).}
\label{fig:mutual_referral}
\end{figure*}

Figure \ref{fig:mutual_referral} presents a visualization of five Brazilian states that have different characteristics not only in terms of population, but also in per capita income and healthcare services availability. S\~ao Paulo and Rio de Janeiro have the larger population numbers, per capita income, and healthcare services offerings. Dots represent physicians, dot size and color are related to node degree, larger/darker dots represent degrees of higher value. The health insurance company has an important presence in S\~ao Paulo and Rio de Janeiro, which can be observed looking at the top 20 physicians in these states. Because of the number of physicians is larger in these states, patients have more options to choose, but they usually choose physicians considering which are the best ones (i.e., by reputation) and these physicians usually work with a very particular group of other physicians (mainly associated with regions and providers). 

On the other hand, in other states such as Bahia, Pernambuco and Distrito Federal, the presence of the health insurance company is sparser; there are less physicians that work with this company, meaning less options for patients and in consequence more evenly distributed degrees, which can also be identified in the density values of the network (see Figure \ref{fig:mutual_referral}). 

We observe an unusual case in Distrito Federal, where two pairs of heavily connected physicians have the median of days between two consultations equal to zero days, meaning that half or more patients consulted both physicians in the same day. This could represent physicians that work in the same healthcare provider, for example a cardiologist that executes an electrocardiogram and another cardiologist that analyzes the result, working together in same clinic. However, in both cases we did not have the physicians specialty information. Actually, most of the physicians that are strongly connected with others do not specify their specialty, which hinders any detailed inference regarding the context in which the connection occurred. 

Table \ref{tab:CRM--CRM-mis-ranking} shows the strongest pairs of physicians in the studied database. Except the top pairs from Mato Grosso do Sul, P$_{MS}$028 and P$_{MS}$027, all the other physicians are from the 5 states represented in Figure \ref{fig:mutual_referral}. Physicians P$_{MS}$028 and P$_{MS}$027 have 205 same-patient consultations (first with P$_{MS}$028 and then with P$_{MS}$027) and 196 consultations (first P$_{MS}$027 and then P$_{MS}$028). None informed their specialty, but with this information, subject matter experts can analyze in detail the highlighted relationships, identifying groups of physicians that work together, misuse health insurance resources by forcing unnecessary referrals, or cases of referral/counter-referral.

\begin{table}[!bp]
\small
\centering
\begin{tabular}{|l|l|r|r|r|}
\cline{3-4}
\multicolumn{2}{c}{} & \multicolumn{2}{|c|}{\textbf{Consultations}} & \multicolumn{1}{c}{} \\ \hline
$P_1$ & $P_2$ & $P_1 \rightarrow P_2$ & $P_2 \rightarrow P_1$ & $mrs(P_1,P_2)$ \\ \hline
P$_{MS}$028 & P$_{MS}$027 & 205 & 196 & 1.000\\ \hline
P$_{DF}$010 & P$_{DF}$009 & 267 & 108 & 0.551\\ \hline
P$_{SP}$022 & P$_{SP}$021 & 103 & 102 & 0.520\\ \hline
P$_{SP}$139 & P$_{SP}$138 & 92 & 72 & 0.367\\ \hline
P$_{DF}$057 & P$_{DF}$056 & 73 & 71 & 0.362\\ \hline
P$_{SP}$141 & P$_{SP}$140 & 72 & 70 & 0.357\\ \hline
P$_{SP}$139 & P$_{SP}$140 & 73 & 66 & 0.337\\ \hline
P$_{SP}$024 & P$_{SP}$023 & 92 & 63 & 0.321\\ \hline
P$_{SP}$143 & P$_{SP}$142 & 73 & 63 & 0.321\\ \hline
P$_{SP}$145 & P$_{SP}$144 & 70 & 62 & 0.316\\ \hline
\end{tabular}
\caption{The top 10 mutual referral scores in the database; 
physicians IDs are anonymized and state codes are used to allow interpretation considering different Brazilian states.}
\label{tab:CRM--CRM-mis-ranking}
\vspace*{-5mm}
\end{table}


Considering the whole network, analysis of those pairs of physicians with highest mutual referral score and with informed specialties revealed the following common specialties referrals:
Ophthalmologist $\leftrightarrow$ ophthalmologist;
Cardiologists $\leftrightarrow$ cardiologists/vascular surgery;
Acupuncture/pediatric $\leftrightarrow$ allergist;
Cardiology $\leftrightarrow$ hematology/clinical pathology;
Dermatology $\leftrightarrow$ cardiology.

\begin{figure*}[hbtp]
\centering
\begin{subfigure}{.33\textwidth}
 \centering
 \includegraphics[width=0.95\linewidth]{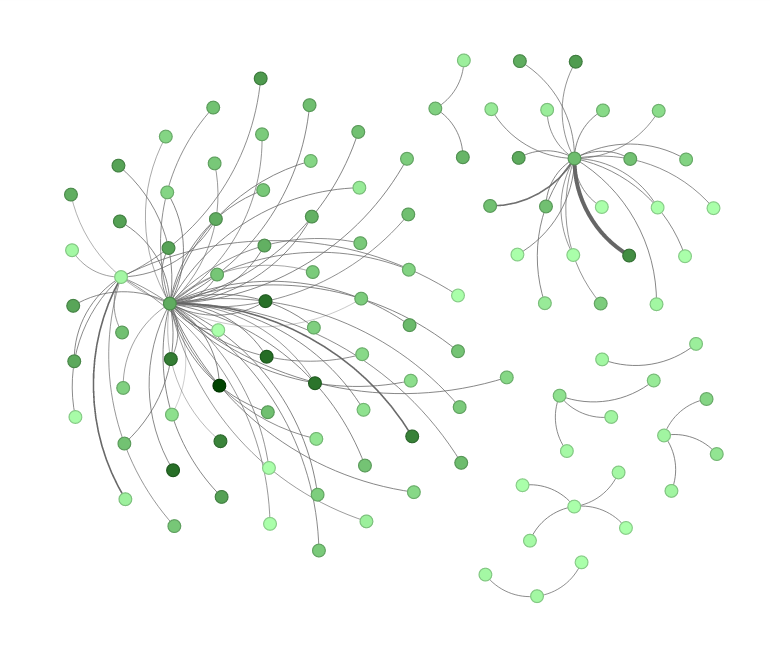}
  \caption{Allergist}
  \label{fig:sub1}
\end{subfigure}
\begin{subfigure}{.33\textwidth}
  \centering
  \includegraphics[width=0.95\linewidth]{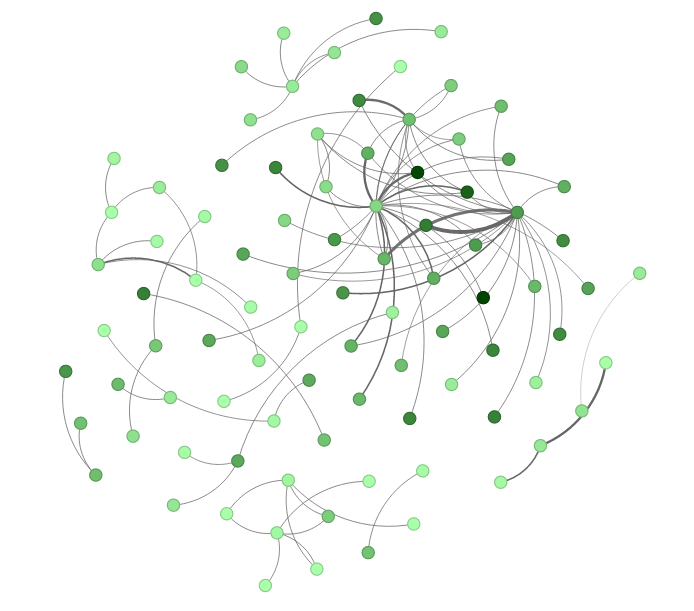} 
  \caption{Cardiologist}
  \label{fig:sub2}
\end{subfigure}
\begin{subfigure}{.33\textwidth}
  \centering
  \includegraphics[width=0.95\linewidth]{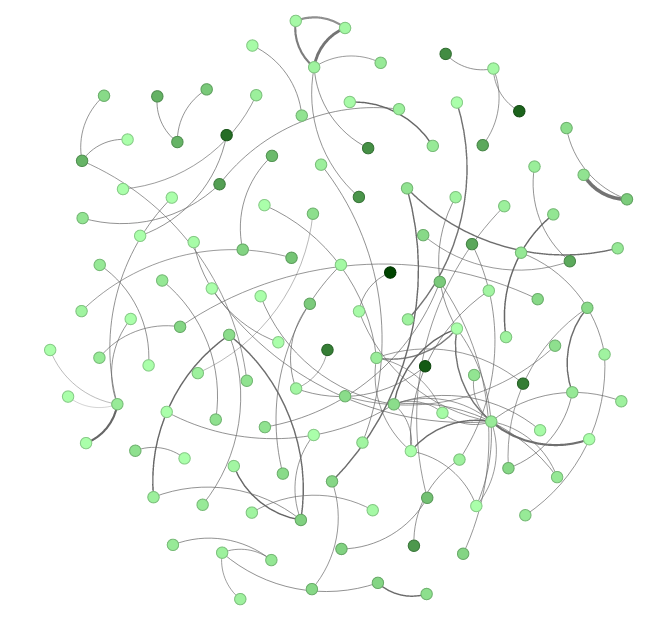}
  \caption{Ophthalmologist}
  \label{fig:sub3}
\end{subfigure}
\caption{Top 100 mutual referral scores by the following specialties:  cardiologist and ophthalmologist.}
\label{fig:allnet}
\end{figure*}

In Figure \ref{fig:allnet} we show the network for the top 100 mutual referral scores by the following specialties: allergist, cardiologist and ophthalmologist. We can see that the behavior is different for each specialty. For allergists (Figure \ref{fig:allnet} (a)) there are two main physicians connected with several other physicians from different specialties, most of them pediatricians. Thus, it is possible to infer that these two allergists are recommended by most of their colleagues. For cardiologists (Figure \ref{fig:allnet} (b)) there are two physicians that are strongly connected with several other physicians and nine other cardiologists that have a small number of other physicians connected to them. Finally, the network is different for ophthalmologists, we have almost no other physician strongly connected to multiple physicians. They work mostly in pairs (usually the other ophthalmologist is a surgeon, or a peer responsible of running certain types of exams).  In the case of ophthalmologists, they tend to be very specialized (e.g., surgeon, retina, cornea, etc.) and so they usually work with a peer.

Lastly, this model suggests that some physicians consider their own connections to indicate patients. Hence, in situations where the health insurance company wants to improve the relationship with physicians in their network, it could consider an approach involving groups of doctors that already act together, increasing the involvement of the group as a whole with the health insurance company and its services. More generally, this metric could support decision making processes involving increasing or reducing the network of accredited physicians/service providers.
\subsection{Retention}

The main goal of the retention model is to identify physicians according of their patients' loyalty. To do this, we analyze the physician-patient relationship. Generally speaking, for each patient $i$ there is a number of claims $s_i$ related to a number of physicians $K_i$ that the patient has visited. We focused exclusively on consultations, each claim corresponding to only one consultation. The $s_i$ claims are thus distributed among $P_i$ physicians in such a way that physician $j$ has $s_{ij}$ claims related to a patient $i$. Thus, the total number of claims for patient $i$ is equal to the sum of all the claims $s_{ij}$ across $K_i$ physicians, that is:

\[ s_i = \sum_{j=1}^{P_i} s_{ij} \]

\begin{defn}[Relative Relationship]
  To capture the relative difference between patient-physician relationships, we define a model called \textit{Relative Relationship} $r_{ij} = s_{ij}/s_i$, as a measure to gauge the strength of the relationship between patient $i$ and physician $j$. If the patient has a large number of claims with a given physician, i.e., $s_{ij} \gg 1$, there is a strong relationship between them. But if there is not only a large number of claims with physician $j$, but also most of those claims are with this specific physician, i.e., $r_{ij}=s_{ij}/s_i \approx 1$, then this is an even stronger indication that there is a tight relative relationship between patient $i$ and physician $j$. A consistently high value of $r_{ij}$ for a given physician $j$ across patients could be an indicator that this professional provides high quality service, motivating patients to return over time. 
\end{defn} 

\begin{defn}[Maximum Relative Relationship]
We consider \textit{Maximum Relative Relationship} $r_{ij}^{max}$ as the highest value of the relative relationship $r_{ij}$ of patient $i$ with physician $j$. Maximum Relative Relationship singles out the most prominent physician for patient $i$, the rationale being to identify physicians that can be in the category of high quality service. This metric is associated to patient-physician pairs and it is particularly significant as it approaches 1 and, simultaneously, the patient is connected to a large number of physicians (i.e., $P_i \gg 1$). 
\end{defn}

To capture a physician's retention capacity in a single metric, we consider a function proportional to the Maximum Relative Relationship $r_{ij}^{max}$ as well as proportional to the number of physicians that the patient consulted with. If the patient consulted with multiple physicians, a
higher number of $r_{ij}^{max}$ for one of them is significant. A physician showing patients that have consistently higher values of $r_{ij}^{max}$ is indicative of the physician's capacity of retaining patients. We average all patients for a given physician and normalize by the maximum value to keep the metric in the unit range, allowing for straightforward comparison.

\begin{figure}[hbt]
\centering
\includegraphics[width=0.45\textwidth]{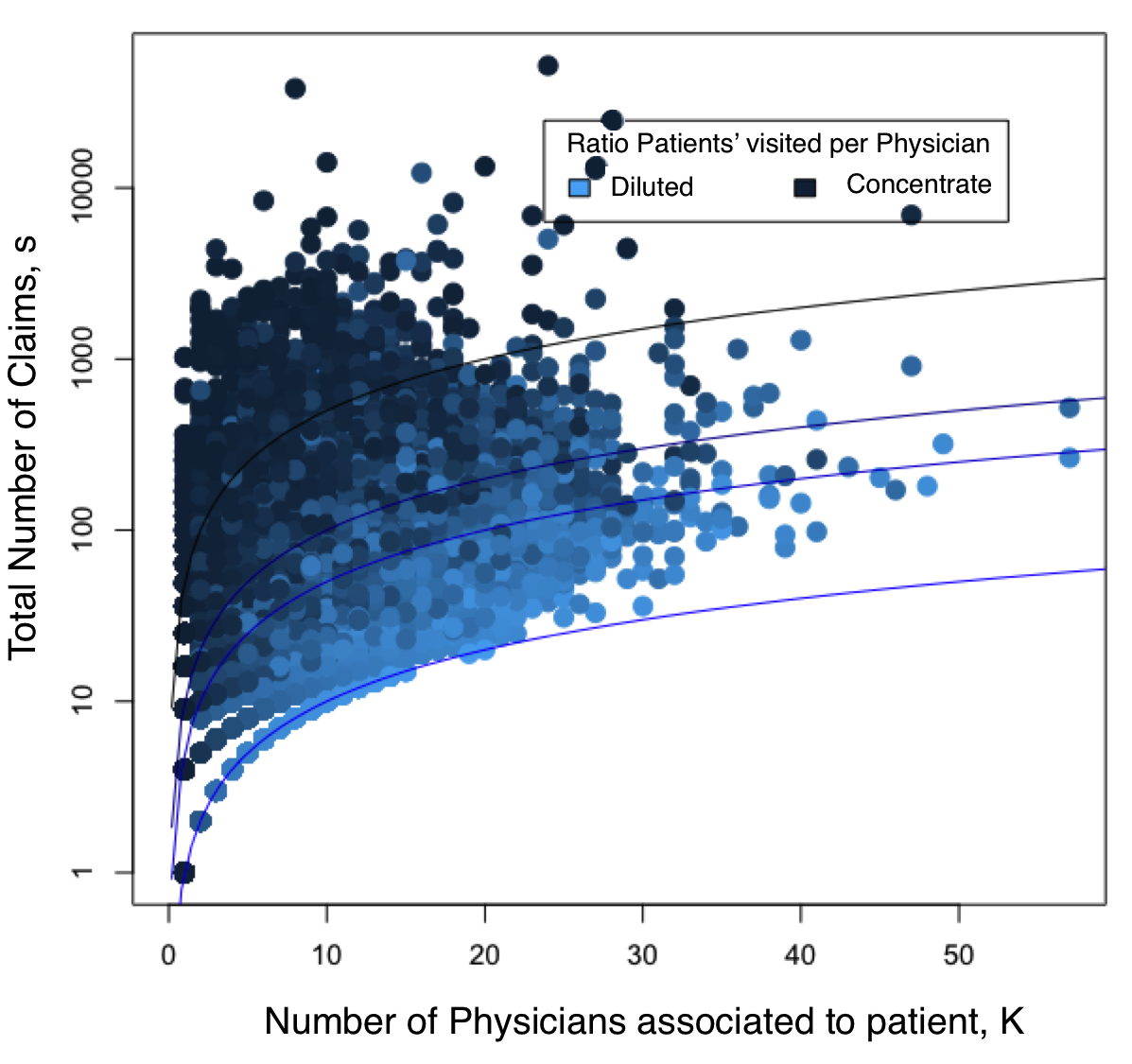}
\caption{Relationship between number of physicians and total number of claims of a patient (each dot represents a patient). The data correspond to consultations in the state of S\~ao Paulo. 
}
\label{fig:diluidoconcentrado}
\end{figure}

Figure \ref{fig:diluidoconcentrado} shows how the Maximum Relative Relationship metric is distributed across all patients in the state of S\~ao Paulo. In the figure, each dot is a patient, the $x$-axis represents the number of physicians connected to the patient, which we denote as $K$, and the (logarithmic) $y$-axis represents the total number of claims of the patient, denoted as $s$. Color intensity is proportional to the value of $r_{ij}^{max}$, so the darker the intensity, the closer to one, which means that the patient has most claims mainly with the corresponding physician. Solid lines represent proportional values of $s$ vs. $P$, i.e. $s = c P$, with $c$ equal to 1, 5, 10, 50. This means that dots above the upper solid line have more than 50 times claims than the corresponding number of visited physicians. 

Our model allows for the definition of parameters to select physicians that have, at the same time, higher values of Maximum Relative Relationship $r_{ij}^{max}$, total number of physicians $P$ and total number of claims $s$, meaning that the physician has patients that have several claims and also physicians built a strong relationship with the patient. In Figure \ref{fig:diluidoconcentrado}, this correspond to the darker points above the black upper line.

This type of information represents a summary of the relationship patient-physician, where is possible to quantify which physician has in general a more loyal relationship from patients, inferred by their returning behavior. 

\begin{table}[hbtp]
\small
\centering
\begin{tabular}{|l|l|l|l|l|l|}
\cline{2-6}
\multicolumn{1}{c}{} & \multicolumn{3}{|c|}{\textbf{2013}} & \multicolumn{2}{|c|}{\textbf{2014}} \\ \hline
\textbf{Physician} &  \textbf{Q2} &  \textbf{Q3} &  \textbf{Q4} &  \textbf{Q1} & \textbf{Q2}  \\ \hline
P$_{BA}$262 & 1 & 1 & 1 & 2 & 2 \\ \hline
P$_{BA}$051 & 2 & 5 & - & - & 4 \\ \hline
P$_{SP}$263 & 3 & - & - & - & - \\ \hline
P$_{SP}$264 & 4 & 2 & 2 & 1 & 1 \\ \hline
P$_{SP}$265 & 5 & 10 & 4 & - & - \\ \hline
P$_{BA}$266 & 6 & - & - & - & - \\ \hline
P$_{SP}$142 & 7 & - & - & - & - \\ \hline
P$_{SP}$267 & 8 & 9 & - & 7 & - \\ \hline
P$_{MG}$268 & 9 & - & 7 & - & - \\ \hline
P$_{DF}$234 & 10 & - & - & - & - \\ \hline
P$_{SP}$269 & - & 3 & - & - & 7 \\ \hline
P$_{DF}$219 & - & 4 & 3 & - & - \\ \hline
P$_{SP}$270 & - & 6 & - & - & - \\ \hline
P$_{DF}$233 & - & 7 & 5 & - & - \\ \hline
P$_{MS}$271 & - & 8 & - & - & - \\ \hline
P$_{BA}$272 & - & - & 6 & 4 & 8 \\ \hline
P$_{SP}$273 & - & - & 8 & - & - \\ \hline
P$_{SP}$274 & - & - & 9 & - & - \\ \hline
P$_{SP}$275 & - & - & 10 & - & -\\ \hline
P$_{SP}$276 & - & - & - & 3 & - \\ \hline
P$_{SP}$277 & - & - & - & 5 & 3 \\ \hline
P$_{DF}$278 & - & - & - & 6 & - \\ \hline
P$_{SP}$279 & - & - & - & 8 & - \\ \hline
P$_{MG}$238 & - & - & - & 9 & 6 \\ \hline
P$_{SP}$280 & - & - & - & 10 & 9 \\ \hline
P$_{DF}$281 & - & - & - & - & 5 \\ \hline
P$_{SP}$282 & - & - & - & - & 10 \\ \hline
\end{tabular}
\caption{Top 10 physicians with higher values for retention for 5 consecutive quarters. We observe that S\~ao Paulo is not the dominant state as opposed to the centrality metrics (See Section \ref{sec:centrality}). There is some stability in the sense that many physicians appear in the top 10 in more than one quarter. 
}
\label{tab:RL3_comparison}
\vspace*{-5mm}
\end{table}

Table \ref{tab:RL3_comparison} shows a comparison of the top 10 physicians from the whole database through five consecutive quarters. We can see that the retention is stable in time since several physicians appear in consecutive quarters systematically.

\subsection{Physician Centrality}
\label{sec:centrality}

In this part of the analysis we focus on analyzing the \emph{influence} of the physician relative to
other physicians in the network. The goal is to understand the physicians activity regarding other
physicians, specially from the point of view of the activity that is seen by the health insurance
company, i.e., the data present in the claims database. The rationale tested here is that influential (more central)
physicians will have a larger effect in the topology of the network than physicians that are in the network's periphery, for instance, an influential physician leaving the network may induce close peers to leave as well ~\cite{centrality2005,borgatti2005centrality}.

To capture the importance of each physician, the following four centrality measures were considered:
\textbf{\textit{Degree}}: The importance of the physician is proportional to the number of patients shared with other physicians. The higher the number of patients shared with other physicians, the more central the physician is regarded. \textbf{\textit{Eigenvalue}}: The higher the number of shared patients with other \emph{important} physicians, the more important the physician is considered in the network. If the physician shares a large number of patients with physicians not considered as important, the physician is not necessarily considered influential \cite{ruhnau2000eigenvector}. \textbf{\textit{Betweenness}}: Takes into account that physicians can be influential to other physicians when they are \textit{central} in the network. It is possible to define a metric that captures this proximity effect by means of the \emph{betweenness coefficient}~\cite{betweenness2004}, which considers the shortest path a patient would take from a given physician to another one, if they make direct referrals. Through this metric, the higher the potential to connect with other doctors, the more central the physician. \textbf{\textit{Closeness}}: Physicians can also be ranked by the amount of other physicians close to each other, where proximity in this case refers to the number of links between physicians. This can me modeled by means of the \emph{closeness coefficient}~\cite{rochat2009closeness}. Thus, it is possible to compute a distance between physicians by considering the (smallest) number of physicians that are needed to connect any two physicians in the network.

These four metrics were computed for the same network, including all Brazilian states, considering also how they change over time by dividing the complete time period in quarters.  
One important characteristic is the possibility of following the physicians evolution related to these metrics. In table \ref{tab:RL2-rank-all} we show the top 10 physicians using the eigenvalue centrality measure, considering all states.  As we can see, some physicians have a stable behavior over time, meaning that they are occupying closer positions, for example, physician P$_{SP}$153 holds the top position during most quarters except in Q4 of 2013 when he/she moved to the second position. In other cases, a physician has an ascendant or descendant ranking, which can indicate a change in the physician's area or patients' behavior that might be of interest for the health insurance company. 

The resulting ranking depends on the used metric, but there is still a high degree of correlation among the metrics, as expected. For instance, for Q2 of 2014, 14 physicians from top 20 using eigenvalue as a metric are in the same position given by the other metrics, showing that the four centrality measures are quite consistent and indicate the importance of the physicians for the network in a similar way. In general, we have a metric concordance between 10\% and 20\% of top 100 physicians, which is remarkable considering that there are more than 8,000 physicians.
\begin{table}[!bp]
\small
\centering
\begin{tabular}{|l|c|c|c|c|c|}
\cline{2-6}
\multicolumn{1}{c}{} & \multicolumn{3}{|c|}{\textbf{2013}} & \multicolumn{2}{|c|}{\textbf{2014}}  \\ \hline
\textbf{Physician} &  \textbf{Q2} &  \textbf{Q3} &  \textbf{Q4} &  \textbf{Q1} &  \textbf{Q2} \\ \hline 
P$_{SP}$153 & 1 & 1 & 2 & 1 & 1 \\ \hline 
P$_{SP}$154 & 2 & 3 & 1 & 2 & 2 \\ \hline 
P$_{SP}$164 & 3 & 4 & 7 & 4 & 5 \\ \hline 
P$_{SP}$142 & 4 & 5 & 3 & 3 & 3 \\ \hline 
P$_{SP}$242 & 5 & 10 & 10 & 14 & 12 \\ \hline 
P$_{SP}$243 & 6 & 8 & 8 & 6 & 8 \\ \hline 
P$_{SP}$140 & 7 & 2 & 4 & 5 & 4 \\ \hline 
P$_{SP}$148 & 8 & 7 & 5 & 13 & 11 \\ \hline 
P$_{SP}$149 & 9 & 12 & 14 & 12 & 16 \\ \hline 
P$_{SP}$244 & 10 & 13 & 12 & 15 & 10 \\ \hline 
P$_{SP}$246 & 12 & 6 & 11 & 7 & 15 \\ \hline 
P$_{SP}$139 & 13 & - & 17 & 18 & 7 \\ \hline 
P$_{SP}$247 & 14 & 20 & 19 & - & 9 \\ \hline 
P$_{SP}$248 & 15 & 9 & 6 & 9 & 18 \\ \hline 
P$_{SP}$143 & 19 & - & - & 10 & 13 \\ \hline 
P$_{SP}$252 & - & 14 & 9 & 8 & 6 \\ \hline 
\end{tabular}
\caption{Temporal evolution by quarters of the top 10 physicians with highest centrality measures in the physician-physician network, considering all Brazilian states.
}
\label{tab:RL2-rank-all}
\vspace*{-5mm}
\end{table}
The analysis of highest ranking physicians by eigenvalues revealed that the group is almost exclusively composed of the same group physicians. In addition, four of these physicians are associated geographically, being P$_{SP}$140 (Otolaryngologist), P$_{SP}$164 (Dermatologist), P$_{SP}$154 (Endocrinologist), P$_{SP}$142 (Hematologist), suggesting that location is a highly relevant factor for such analysis. The fifth physician, P$_{SP}$153, is a well known cardiologist from S\~ao Paulo, SP.

\begin{figure}[htbp]
\centering
\includegraphics[width=0.45\textwidth]{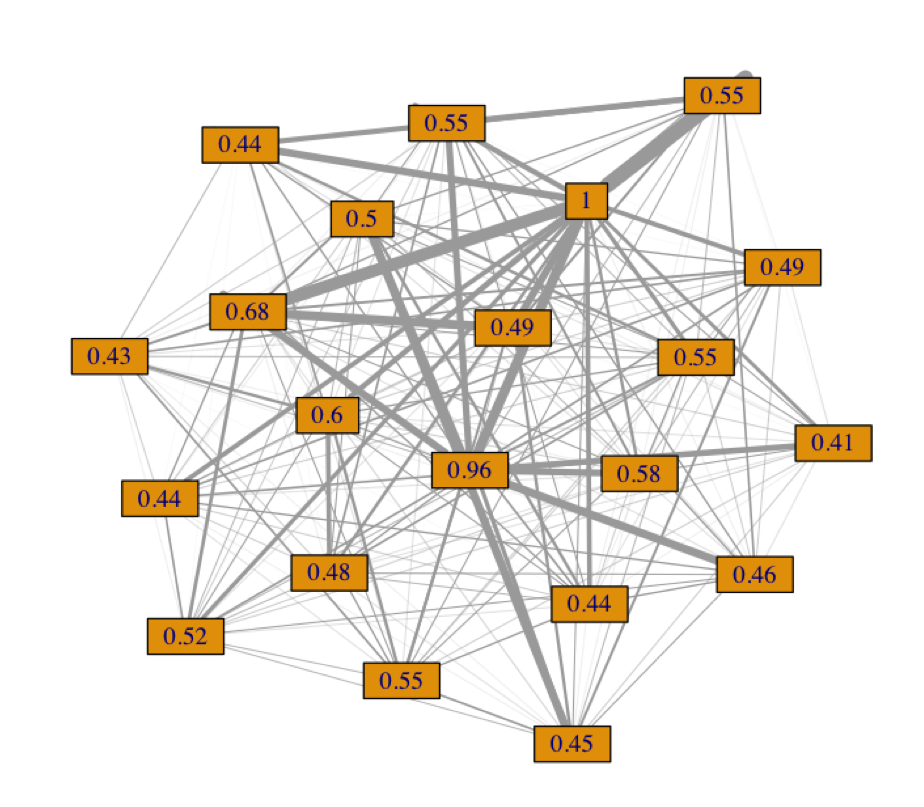}
\caption{Subset of 21 physicians with highest network importance for Q2 2014. The label inside the node represents the physicians' eigenvalue. The width of the edges is proportional to the number of shared patients. 
}
\label{fig:core}
\end{figure}

In order to analyze the connectivity among physicians, the network density was computed, i.e. that ratio of the number of \textit{existing} links to the total number of \textit{possible} links. For the whole network 
the density was 0.001, for the top 100 physicians the density was 0.6 and for the top 40 physicians the density was 0.9. Figure \ref{fig:core} shows a representation of the network of physicians with highest eigenvalue for Q2 in 2014 considering all Brazilian states. The subset is composed of 21 physicians with density 0.933, which means that they are almost all connected to each other. 

\section{Conclusion} \label{sec:conc}

In this work we have shown how social networking techniques can be applied to the analysis of health
insurance claims data, mainly by mapping physicians using shared patients as a proxy for a relationship between them. The resulting models provided useful insights to the health insurance company we partnered
with. The framework improved the understanding of important characteristics both from
physicians and from patients involved in consultations patterns. Those insights can have multiple
business applications such as to detect frauds or to improve the relationship between the health insurance
company and important physicians for the company (e.g., physicians connected with the core of the
network, with high mutual referral with well-known physicians, and with high retention).


We demonstrated that we can obtain useful insights from claim data if we model as a social network supporting a structural analysis as
opposed to traditional transactional approaches. Moreover, our results point out that the complex
physician-physician network derived from the health insurance claims database has characteristics
similar to a social network, what opens up multiple paths for this research to follow, in particular, 
considering theories and techniques from Social Network Analysis.

The data analysis and the value of the models for the business of the health insurance company we
partnered with were validated by multiple meetings with specialists from our lab and from the company,
including physicians, process analysts, and information technology specialists. Those interactions
with subject matter experts were fundamental for the identification of the most important cases and scenarios to be
considered (e.g., cases where multiple physicians used the same physician ID when submitting a claim
to the company) and for the creation of models to support decision making processes and with real business value.



\bibliographystyle{ACM-Reference-Format}
\bibliography{paper}

\end{document}